\documentclass[pra,aps,showpacs,twocolumn]{revtex4}
\usepackage{amsmath}
\usepackage{amsfonts}
\usepackage{amssymb}
\usepackage{dsfont}
\usepackage{mathrsfs}
\usepackage{epsfig}
\usepackage{bm}
\usepackage{amsbsy}
\usepackage{color}
\usepackage{epstopdf}
\usepackage{txfonts}

\begin{document}


\title{Quantum Model for a Periodically Driven Selectivity Filter in K$^{+}$ Ion Channel}

\author{A. A. Cifuentes}
\affiliation{Centro de Ci\^encias Naturais e Humanas, Universidade Federal do ABC, 09210-170, Santo Andr\'e, S\~ao Paulo, Brazil}\author{F. L. Semi\~ao}
\affiliation{Centro de Ci\^encias Naturais e Humanas, Universidade Federal do ABC, 09210-170, Santo Andr\'e, S\~ao Paulo, Brazil}
\begin{abstract}
In this work, we present a quantum transport model for the selectivity filter in the KcsA potassium ion channel. This model is fully consistent with the fact that two conduction pathways are involved in the translocation of ions thorough the filter, and we show that the presence of a second path may actually bring advantages for the filter as a result of quantum interference. To highlight interferences and resonances in the model, we consider the selectivity filter to be driven by a controlled time-dependent external field which changes the free energy scenario and consequently the conduction of the ions. In particular, we demonstrate that the two-pathway conduction mechanism is more advantageous for the filter when dephasing in the transient configurations is lower than in the main configurations. As a matter of fact, K$^+$ ions in the main configurations are highly coordinated by oxygen atoms of the filter backbone and this increases noise.  Moreover, we also show that, for a wide range of dephasing rates 
and driving 
frequencies, the two-pathway conduction used by the filter leads indeed to higher ionic currents when compared with the single path model. 
\end{abstract}
\pacs{03.65.Yz,	
87.15.A-, 
05.60.Gg,	
}

\maketitle
\section{Introduction}
Quantum biology is an emerging field of research which aims at investigating the possibility of a functional role for quantum mechanics or coherent quantum effects in biological systems. Notably, much of the work made in the last years have been related to transport, being the Fenna-Matthews-Olson (FMO) complex an important example \cite{FMO}. This pigment-protein complex is present in green sulfur bacteria and its function is to channel excitons from the chlorosome antenna complex, where light is harvested, to the reaction center which execute the primary energy conversion reactions of photosynthesis. Much of the attention given to this complex, and to quantum biology, in general, arose from the experimental observation of long-lived oscillatory features using ultrafast 2-D spectroscopy \cite{exp}. Such oscillations were interpreted as evidence for the presence of long-lived electronic coherence, something not trivial given the complexity of these open systems. After these experimental observations, many 
efforts have been made to explain the origin of those coherences \cite{origin}, and also to investigate their possible relation with entanglement and other quantum related effects \cite{qf}. 

Due to its particular features and efficiency, ion channels constitute another biological system where non trivial quantum effects may appear and be functional \cite{original}. These channels are transmembrane proteins and they have an important role in the production of electric signals in biological systems \cite{book}. Their structure gives rise to a selectivity filter which is a very narrow channel which catalyses the dehydration, transfer, and rehydration of the ions in a very efficient way, achieving a flux of about $10^8$ ions per second \cite{ef}. Throughly crystallographic studies and free energy computer simulations showed, more than ten years ago, that ion translocation in the filter involves transitions between two main states, and that these transitions occur through two physically distinct pathways of conduction \cite{Cabral,Berne}. These pathways involve either two or three K$^+$ ions occupying the selectivity filter. Details about the experiment and the simulations demonstrating 
that ion translocation in the channel unmistakably follows two distinct paths can be found in Methods section of \cite{Cabral}.

In this work, we consider a quantum model with two conduction pathways in accordance with experimental results and simulations in potassium channels \cite{Cabral,Berne}. Consequently, the transport in this system constitutes a two-path problem where quantum superposition effects coming from the \textit{competing paths} of conduction play a decisive role. Since we treat here the ionic current in the filter, the results predicted in this work can in principle be experimentally accessed with physiological techniques \cite{original}. This requires the filter to be driven by a periodic time dependent electric field which is also included in our analyses. In the following, we present the basic elements of the model and the inclusion of the driving field. We then study the ionic conduction or current, highlighting the possible advantages of having two and not just one conduction path. In particular, we study the role of having non uniform dephasing in the topology, given that this is the most likely physical 
picture in the selectivity filter.
\section{Modeling Configurations and paths in the channel}
In the systematically studied KcsA potassium channel from soil bacteria \textit{Streptomyces lividians}, whose structure is very well known \cite{struc}, K$^+$ ions loose their hydrating water molecules to enter the selectivity filter and carbonyl oxygen atoms in its backbone replace the water molecules. This allows the formation of a series of coordination shells through which the K$^+$ ions can move. Qualitatively distinguishable configurations of ions and water molecules in the selectivity filter correspond to different configurations which will be represented here as two-level systems. To be more specific, if a site $k$ is populated i.e., in state $|1\rangle_k$, this configuration is active. Otherwise, if it is in state $|0\rangle_k$, this configuration is not active or not participating in the ionic conduction. In a classical hoping mechanism, we would never find superpositions in one configuration (being \textit{and} not being used) or entanglement between different configurations (different sites). 
Such quantum coherent events lead to resonances in the ionic current which might be measured directly using physiological techniques \cite{original}. 

\begin{figure}[hbt]
\centering\includegraphics[scale=0.1]{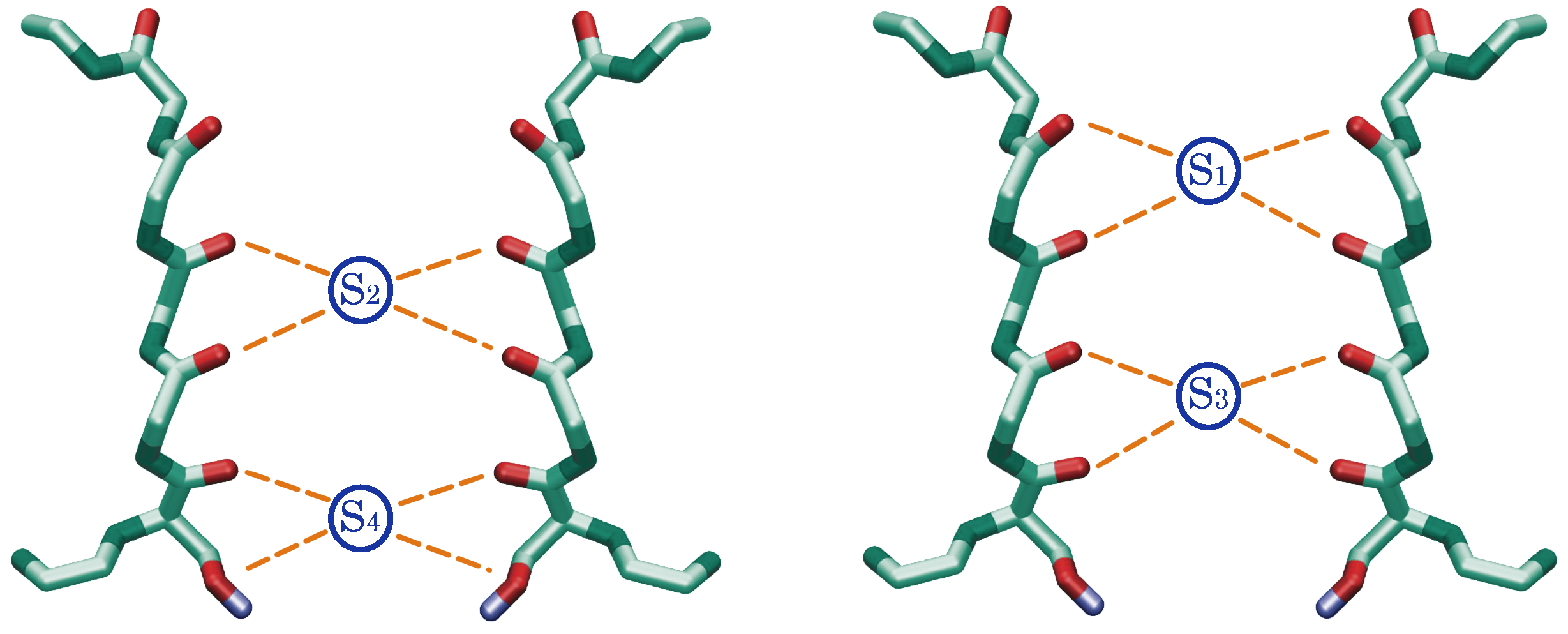}
\caption{In the selectivity filter, ions move outwards the cell in the presence of negative oxygen atoms (red segments) of the carbonyl groups in the lateral backbones. Just two of the four backbones are shown. The concerted motion involves alternating  potassium ions K$^+$ and water molecules H$_2$O (not shown). $S_1-S_4$ are binding sites which are numbered according to convention that numbers decrease when the ion proceeds from intra- to extracellular space.} 
\label{s}   
\end{figure}

Ion translocation in the KcsA consists of K$^+$ ions that proceed along the pore
axis of the selectivity filter in a single fashion with water molecules intercalating
them. This gives rise to two main configurations, usually denoted as {\bf{1-3}} and {\bf{2-4}}, representing the position of a pair of K$^+$ ions in the selectivity filter as depicted in Figure \ref{s}. The smaller these numbers are, the closer to the extracellular side the ions are. For the sake of simplicity in notation, we will denote here {\bf{2-4}} by $s$ (source) and {\bf{1-3}} by $d$ (drain), indicating that we will assume that {\bf{2-4}} is in part driven incoherently by interactions of intracellular K$^+$ ions with the carbonyl oxygens in the entrance of the channel, and {\bf{1-3}} can decay incoherently to another configuration culminating with an ion leaving the cell. These processes can be described by Lindblad superoperators in the form \cite{Lind}
\begin{eqnarray}\label{l1}
\mathcal{L}_{s}(\rho)&=&\Gamma_{s}\left(-\left\{\sigma_{s}^-\sigma_{s}^+,\rho\right\}+2\sigma_{s}^+\rho\sigma_{s}^-\right),\\ \label{l2}
\mathcal{L}_{d}(\rho)&=&\Gamma_{d}\left(-\left\{\sigma_{d}^+\sigma_{d}^-,\rho\right\}+2\sigma_{d}^-\rho\sigma_{d}^+\right),
\end{eqnarray}
where $\mathcal{L}_{s}(\rho)$ ($\mathcal{L}_{d}(\rho)$) causes incoherent pump (decay) in the source (drain), $\Gamma_{s}$ ($\Gamma_{d}$) is the incoherent pump (decay) rate for the source (drain), and $\sigma_k^+$ ($\sigma_k^-$) is the two-level raising and lowering operator which create (destroy) excitations in site $k=s,d$. Hereafter, $\left\{\star,\rho\right\}$ denotes the anticommutator $\left\{\star,\rho\right\}\equiv\star\,\rho+\rho\,\star$ and $\rho$ is the density operator for the four sites.

\begin{figure}[hbt]
\includegraphics[scale=0.11]{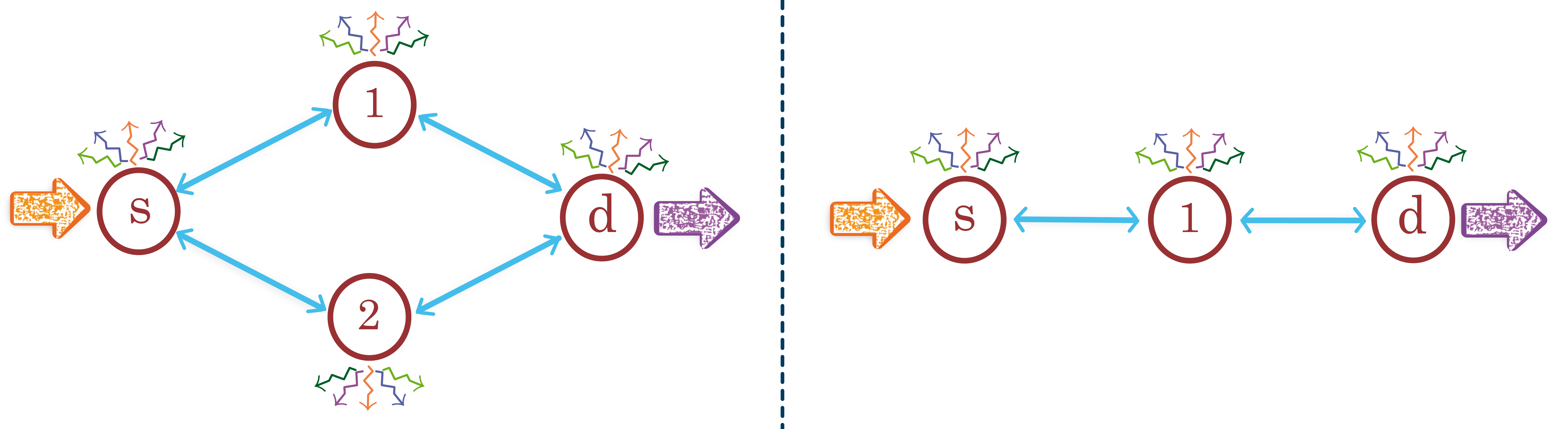}
\caption{(left) Two-path topology where the source $s$ is incoherently driven and excitations hop thorough the network until it is incoherently dissipated through the drain $d$. (right) Single-path topology. All sites are subjected to dephasing (wavy arrows). The selectivity filter in potassium ion channels uses a two-path configuration for changing between $s$ and $d$.} 
\label{top}   
\end{figure}

There are two optimal pathways connecting $s$ and $d$ \cite{Cabral,Berne}, and they will be denoted here by numbers $1$ and $2$, with no risk of confusing them with the main configurations since these are now denoted as $s$ and $d$. Figure \ref{top} depicts this two-path network on the left panel. On the right panel, we present a linear single-path chain which will serve as a benchmark to test for a possible quantum advantage of the two pathway conduction. Along pathway $1$, an ion first
approaches the intracellular entrance to the selectivity filter, and
then pushes the two ions in the filter, causing the outermost ion to leave the channel into the extracellular side. This is known as \textit{concentration-dependent path} \cite{Cabral}. In the second optimal pathway, the \textit{concentration-independent path}, the two ions in the
filter move first, leaving a gap in the selectivity filter which will attract an incoming ion from intracellular space. All microscopic elementary steps involved in these transitions can occur reversibly \cite{Berne}. For this reason, we represent the situation by a hopping term in the Hamiltonian ($\hbar=1$),
\begin{eqnarray}\label{hhop}
H_{hop}&=& c(\sigma_s^+\sigma_{1}^-+\sigma_s^-\sigma_{1}^+)+c(\sigma_1^+\sigma_{d}^-+\sigma_1^-\sigma_{d}^+)\nonumber\\
&&+\beta c(\sigma_s^+\sigma_{2}^-+\sigma_s^-\sigma_{2}^+)+\beta c(\sigma_2^+\sigma_{d}^-+\sigma_2^-\sigma_{d}^+),
\end{eqnarray} 
where $c$ is a hopping rate and $\sigma_k^+$ ($\sigma_k^-$) with $k=1,2$ is the two-level raising (lowering) operator for site $k$. For the two (single) pathway topology we take $\beta=1$ $(\beta=0)$. Of course, these configuration changes do not take place at the same rate, but we kept them equal in this treatment for the sake of simplicity. Our main goal is to discuss the general aspects of the problem.

\section{Driving the Channel}
The last section, and especially Hamiltonian (\ref{hhop}), refers to the channel under natural conditions in the cell membrane. In this case, there are only small electric fields due to concentrations of different ions inside and outside the cell and also charged residues of aminoacids forming the filter. Now, we will consider a particular technique which allows one to probe individual channels in the membrane and to subject them to different electric fields and chemical environments. This technique, called patch clamping \cite{patchC}, enables one to subject the ions in the channel to constant and time-dependent potentials due to applied electric fields \cite{original}. Consequently, this changes the free energy scenario which rules the translocation of ions in the filter \cite{Cabral}. Following \cite{original}, we will consider the field to be engineered so that the configurations follow
\begin{eqnarray}\label{hex}
H_{ext}&=&(\Omega_0+\Omega_1\cos\omega t)[\sigma_s^+\sigma_s^-+2(\sigma_1^+\sigma_1^-+\beta\sigma_2^+\sigma_2^-)\nonumber\\ &&+3\sigma_d^+\sigma_d^-],
\end{eqnarray}   
where $\Omega_0$ and $\Omega_1$ are essentially the amplitudes of the dc and ac parts of the field, respectively, and $\omega$ the angular frequency of the ac part. Similar changes in the free energy scenario occur due to long-range
coupling mechanisms in response to a perturbation at a
large distance in the protein. This is what happens, for instance, in ion pumps due to the binding of Adenosine triphosphate (ATP). 

The full Hamiltonian for the periodically driven selectivity filter is then $H=H_{hop}+H_{ext}$. And this does not refer to the channel under natural condition but rather to the channel being probed in the patch clamping setup. Interaction with the environmental degrees of freedom, especially vibrations of the carbonyl groups of the selectivity filter backbone \citep{origin}, surely induces dephasing noise. In this first treatment of the problem, we will assume the simplest model where this noise is local and memoryless, i.e., we will use the following Lindblad superoperator
\begin{eqnarray}\label{deph}
\mathcal{L}_{deph}(\rho)=\sum_{i=s,d,1,2}\gamma_i\left(-\left\{\sigma_i^+\sigma_i^-,\rho\right\}+2\sigma_i^+\sigma_i^-\rho\sigma_i^+\sigma_i^-\right),
\end{eqnarray}
where $\gamma_i$ is a time independent positive dephasing rate. Usually, the drain excitation probability or population is used to quantify transport efficiency in coupled quantum systems \cite{Plenio,Rebentrost,semiao,Chin}. Here, we are interested in the time average of this quantity or current $I$ which reads \cite{original}
\begin{equation}\label{i}
I=\lim_{T\rightarrow\infty}\frac{1}{T}\int_{0}^{T}\,2\,\Gamma_{d}\rho_{d}(t)\,dt,
\end{equation}
where $\rho_{d}(t)$ is the reduced state of the drain which is obtained by tracing out all other sites.  

Both, the concentration dependent and independent paths, appear in the experiments and simulations as pathways connecting site $s$ to site $d$, and then contribute to ionic conduction \cite{Cabral,Berne}. In the quantum regime, these different paths may \emph{compete} leading to interference effects in the observable current $I$. We now present our findings about the main trends followed by this current. In particular, we investigate a possible quantum advantage of having two conduction paths linking $s$ and $d$ by comparing the current $I$ produced with the topologies shown in Figure \ref{top}. Although we are treating transport in the context of a biological system, it is worthwhile noticing that coupled two-level systems also appears in a great variety of physical scenarios including, for example, quantum dots \cite{semiao} and superconducting qubits \cite{sq}. Consequently, the results presented here might be of value for quantum technologies using qubits, where our results might even be promptly 
simulated and experimentally observed \cite{prep}.
\section{Estimation of Parameters and Decoherence in the System}
In this section, we provide a more detailed physical discussion about the choice of parameters used in the simulations presented later on in this work. In particular, we try to predict the order of magnitude of the decoherence rates in order to compare it with estimated frequencies of the system. It is important to remark that our work is based on an \textit{effective model}. The same is done in recent descriptions of the FMO \cite{Plenio,Rebentrost}. In this case, nonlinear spectroscopy techniques gives direct information on coupling constants and frequencies appearing in an effective description based on the occupation of two-level sites. For the ion channels, however, there is not yet an experimental technique which provides information on the parameters used in effective models. Consequently, we have to estimate the order of magnitude of the parameters involved in our model using indirect available experimental data. This is of course not the most suitable scenario to do predictions, but we 
think that this is not a reason to prevent serious investigations which pave the way for advances and motivate the proposition of new experimental techniques to verify or falsify the findings of the models.

The physical constants should be chosen as to fullfill the expected (and measured) current which is transported by the channel under regular conditions, i.e., $10^8$ ions/sec \cite{ef}. To achieve this, we basically follow the reasoning presented in \cite{original}. From (\ref{hhop}) and (\ref{hex}), it follows that $\Omega_0$ is the energy difference between the configurations and $c$ is the hopping rate between them. Let us consider the case $c<<\Omega_0$. In order to active a transfer rate of $10^8$ ions/sec, perturbation theory tell us that we must have $c^2/\Omega_0\approx 10^8$ions/sec. On the other hand, this time dependent transport model presents resonances when $\Omega_0=n\omega$ with $n$ integer and $\omega$ the frequency of the drive which appears in (\ref{hex}). We will set the system to work near theses resonances. By defining the constant $\omega_0=10^8\text{s}^{-1}$, these conditions are fulfilled by choosing $c\approx \omega_0$ and $n>>1$. For this reason, the simulations are 
run using $c=8\omega_0$ and $\Omega_0=256\omega_0$. The incoherent pump and disposal of configurations given by $\Gamma_{s}$ and $\Gamma_{d}$, respectively, depend on the specific conditions under which the experiment will be set. It is natural to think them as a monotonic function of the concentration of ions inside and outside the cell, as supposed in \cite{original}. Numerical simulations show that variations of these constants only limit the total current but not its dependence on other parameters. So, it does no harm to fix these constants to be the same order as the rate $10^8$ ions/sec, i.e, $\Gamma_{s}=2\omega_0$ and  $\Gamma_{d}=\omega_0$. Finally, one have some freedom to set $\Omega_1$ (the drive amplitude) at any desired value because it is an externally controlled parameter. Since the behavior of the current as a function of $\Omega_1$ consists of a sequence of maxima and minima \cite{original}, we will set this amplitude such that one has the first minimum of current. This is achieved with $\
Omega_1=284.92\omega_0$.

Giving the complexity of the system, it is not easy to anticipate the decoherence rate. For this reason, in the first simulation shown in Figure \ref{fig3}, we vary the dephasing rate $\gamma$ over a wide range. Consequently, we can draw conclusions in regimes such as pure quantum transport (small decoherence) and highly classical transport (massive decoherence). In what regime precisely the channel works is a question to be answered experimentally such as happened to some photosyntetic complexes which were shown to keep track of some quantumness due to partial preservation of quantum coherences \cite{exp}. However, it is interesting to see that it is possible, through simplified assumptions, to provide a rough order of magnitude of the dephasing rate. Again, we follow the reasoning originally shown in \cite{original}. One can assume that the dominant form of noise comes from the stretch mode of the carbonyl groups in the ion channel. This naturally changes the width of the wells forming the trap sites, 
causing the frequency or energy of the stable configurations to fluctuate. The simplified model considering just one trap site and one mode of stretch  is then described by the Hamiltonian 
\begin{equation}\label{decmodel}
 H_{t,CO}=\omega_tb^\dag b+\omega_{\rm{CO}}a^\dag a+\lambda b^\dag b(a^\dag+a),
\end{equation}
where $\omega_t$ is the frequency of ion in the trapping well and $\omega_{\rm{CO}}$ is the frequency of the stretch mode. If one consider the transport of ions as a sequence of tunelling events through barriers separating the wells, it is not hard to show that the frequency of motion in each well must be around $\omega_t\approx 10^{12}$s$^{-1}$ in order to obtain a tunneling rate of $10^8$ ions/sec, as shown in \cite{original}.  Concerning the mode, given its typical high frequency $\omega_{\rm{CO}}\approx 10^{14}$s$^{-1}$, only the ground state is appreciably populated in room temperature and the corresponding mean square deviation of the position of the oxygen atoms in the carbonyl groups is of the order of 0.02$\AA$. One can then attempt to numerically find the order of magnitude of the fluctuations in $\omega_t$ induced by oscillations of amplitude 0.02$\AA$. This fluctuation turns out to be about one order of 
magnitude smaller than $\omega_t$ \cite{original}. Consequently, we can roughly consider $\lambda=\omega_t/10$ in (\ref{decmodel}).

In the scope of this simplified model, an initial superposition state of the ion such as $(|0\rangle+|1\rangle)/\sqrt{2}$ times the ground state of the mode, would evolve such that after a time $t_d\approx 10^3\omega_t^{-1}=1$ns, the fidelity with the initial state will drop to $1/e$. This suggests a decoherence rate  about $\gamma\approx 1/t_d=10^9\text{s}^{-1}$. Therefore, in this crude estimation, one obtains that the dephasing rate $\gamma$ is approximatelly one order of magnitude stronger than $\omega_0$. But this is not enough to discard the analyses of regimes where $\gamma$ is slightly bigger or smaller than $\omega_0$.  In fact, we will show in next section that the two-path topology is more efficient than the single path one even for high dephasings. As said before, it is likely that only an experiment will be able to precisely determine $\gamma$. In \cite{original}, it is proposed the use of physiological techniques to experimentally estimate $\gamma$ from measurements of current.

\section{Simulations}
We first analyze the role of dephasing in the transport, especially in the conduction pathways embodied by sites $1$ and $2$.  It is well known that in the selectivity filter, the configurations $s$ and $d$ have K$^+$ ions residing near the center of a box formed by eight carbonyl oxygens, while in the intermediate sites $1$ and $2$ the coordination is reduced to six oxygen atoms, with just four of them provided by the carbonyl groups of the backbone \cite{Cabral}. Since coupling to the stretch mode of the carbonyl groups is expected to be the main cause of decoherence, we expect that the intermediate sites will possess lower decoherence rates. It is then interesting to see whether having less dephasing in the intermediate configurations helps conduction.  

\begin{figure}[hbt]
\includegraphics[scale=0.23]{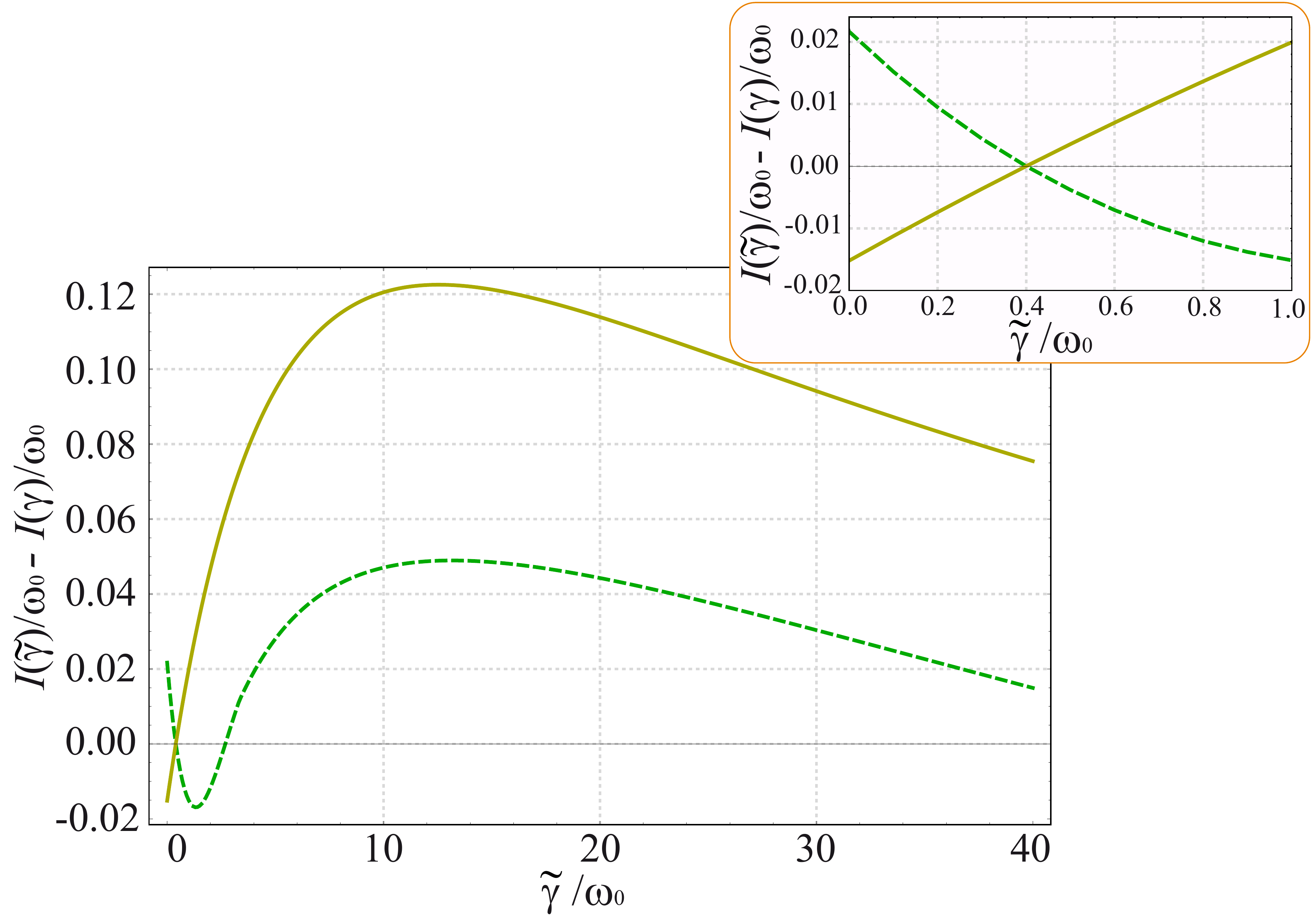}
\caption{Current $I(\tilde{\gamma})$ obtained by fixing dephasing $\gamma=0.4\omega_0$ in sites $s$ and $d$ and varying $\tilde{\gamma}$ which is the dephasing in sites $1$ and $2$. We subtracted $I(\gamma)$ which is the current for equal dephasing in all sites. Solid line corresponds to the single-pathway topology and dashed line to the two-pathway. We used $\omega=4\omega_0$ and $\Omega_1=284.92\omega_0$.} 
\label{fig3}   
\end{figure}

In order to investigate this point, we compare both topologies  considering fixed dephasing in sites $s$ and $d$   ($\gamma_s=\gamma_d=\gamma=0.4\omega_0$), and varying the dephasing $\tilde{\gamma}$ in $1$ and $2$ ($\gamma_1=\gamma_2=\tilde{\gamma}$). The current $I(\tilde{\gamma})$ for this configuration is presented in Figure \ref{fig3}, where we subtracted the current $I(\gamma)$ which corresponds to the case of invariant dephasing. Interestingly enough, it is clear that the two pathway topology benefits from the passing through configurations of reduced dephasing ($\tilde{\gamma}<\gamma=0.4\omega_0$) which is consistent with the fact that the intermediate sites are less coordinated.

\begin{figure}[hbt]
\includegraphics[scale=0.25]{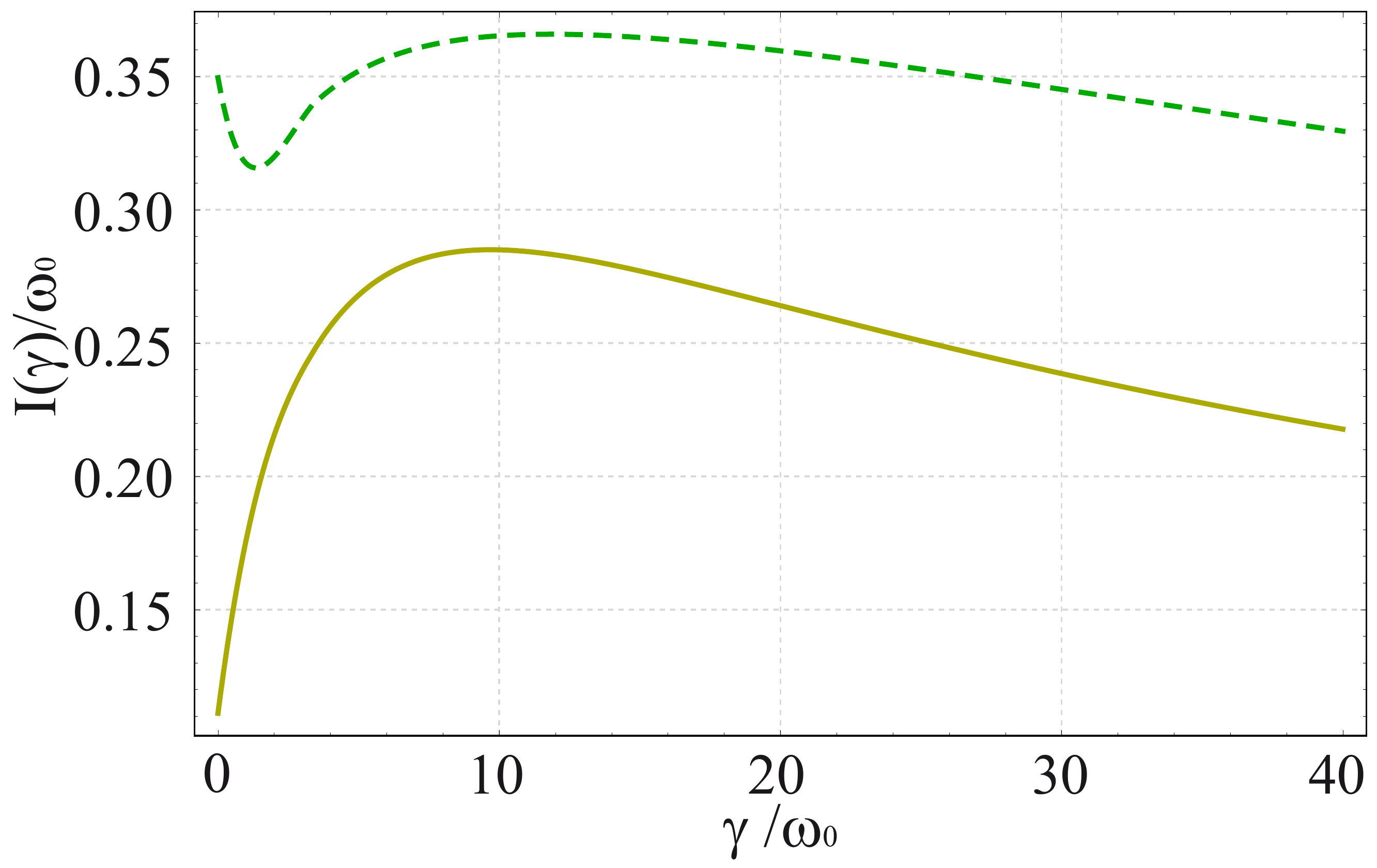}
\includegraphics[scale=0.25]{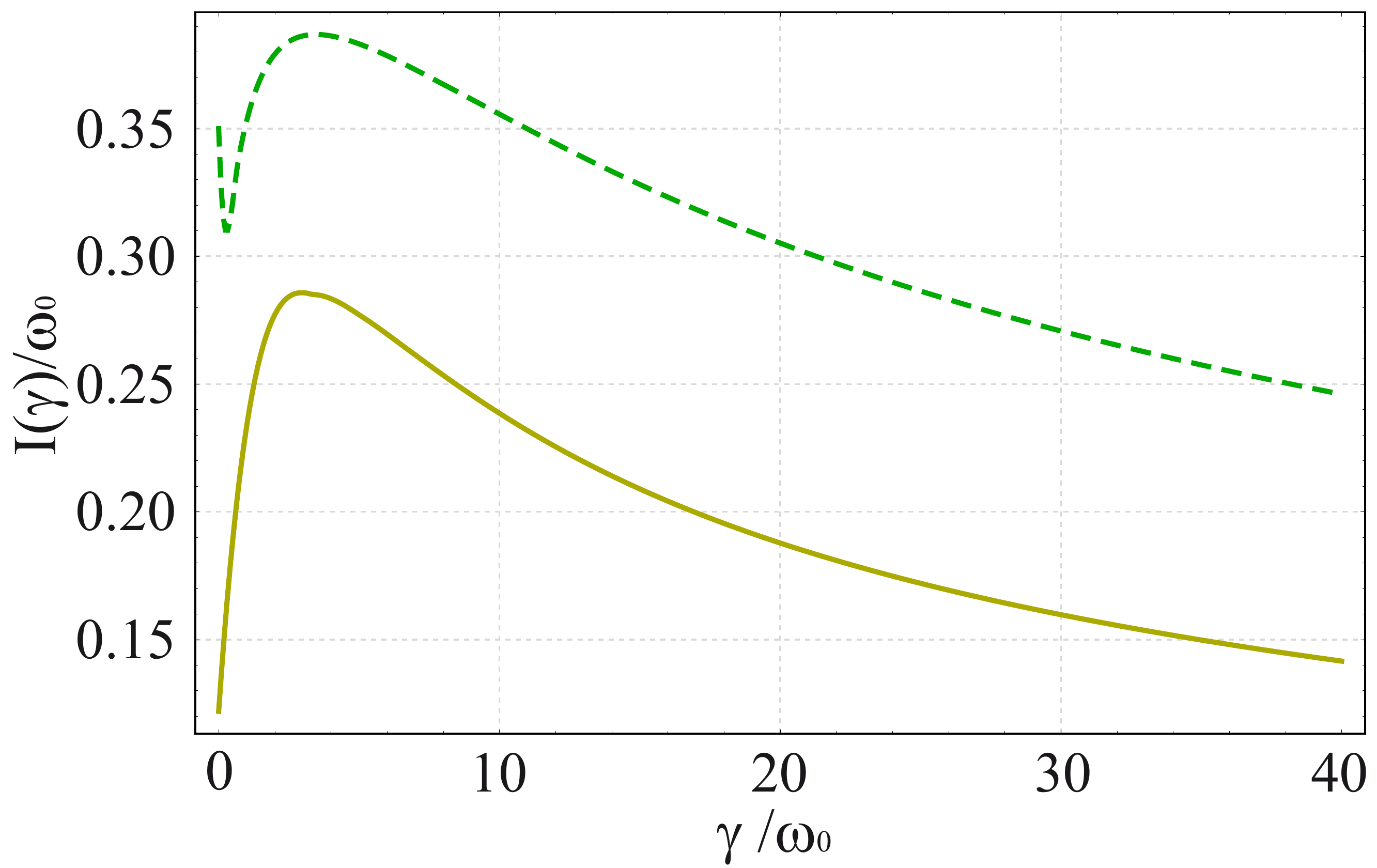}]
\caption{(Above) Current $I(\gamma)$ obtained by fixing $\tilde{\gamma}/\gamma=0.3$ and varying $\gamma$ which is the dephasing in sites $s$ and $d$. The quantity $\tilde{\gamma}$ is the dephasing in the intermediate configurations $1$ and $2$. Solid line corresponds to the single-pathway topology and dashed line to the two-pathway. (Below) The same for $\tilde{\gamma}/\gamma=3$. Solid line corresponds to the single-pathway topology and dashed line to the two-pathway. We used $\omega=4\omega_0$ and $\Omega_1=284.92\omega_0$.} 
\label{fig4}   
\end{figure}

It is import to remark that Figure \ref{fig3} does not allow us to decide which topology is best suited for transport for a given value of $\gamma$. It only allows us to evaluate the advantage of having intermediate sites of lower decoherence rates than the binding sites. In figure \ref{fig4}, we address this point by fixing the ration $\tilde{\gamma}/\gamma$ and varying $\gamma$. We show the behavior of the current in the cases where decoherence in the intermediate sites is lower or stronger than in the binding sites. It is now quite clear that, in respect to different decoherence regimes, the two-pathway topology is more likely to bring transport advantages to the channel compared to the linear topology. 

In general, for a giving value of the driving field frequency $\omega$, models such as the one considered here present resonances when varying the driving field amplitude $\Omega_1$ \cite{original}. In order to gain more information about the advantages of having one or two conduction pathways in the filter, we now study the global maximum of current $I_{{\rm{max}}}(\Omega_1)$ as a function of $\omega$. The result is shown in Figure 5. It is now clear that for most cases the two-path topology can offer advantages with and without dephasing i.e., the range of $\omega$ for which the two-path supersedes the single path is quite wide. The results are actually quite convincing in favor of the two-path topology. Let us take the case $\omega=10\omega_0$, for instance. The current with no dephasing using the two-path topology is more than twice the current observed in the single-path topology. This is a clear evidence of constructive quantum interference arising from competing conduction paths. Even in the presence 
of dephasing, the advantage of the two-path topology at $\omega=10\omega_0$ is much more pronounced than advantage found with the single-path for small $\omega$. Therefore, having two competing paths of conduction gives in general advantages for the filter, and this might have actually been used by channel to help it operate under a real noisy environment. As a final comment, it is interesting to see that about $\omega=12\omega_0$, dephasing helps conduction in the two-path topology. This is a phenomenon called dephasing-assisted transport in the literature \cite{Plenio,Rebentrost}. The same happens for the single-path topology for $\omega$ bigger than about $9\omega_0$.

\begin{figure}[hbt]
\includegraphics[scale=0.25]{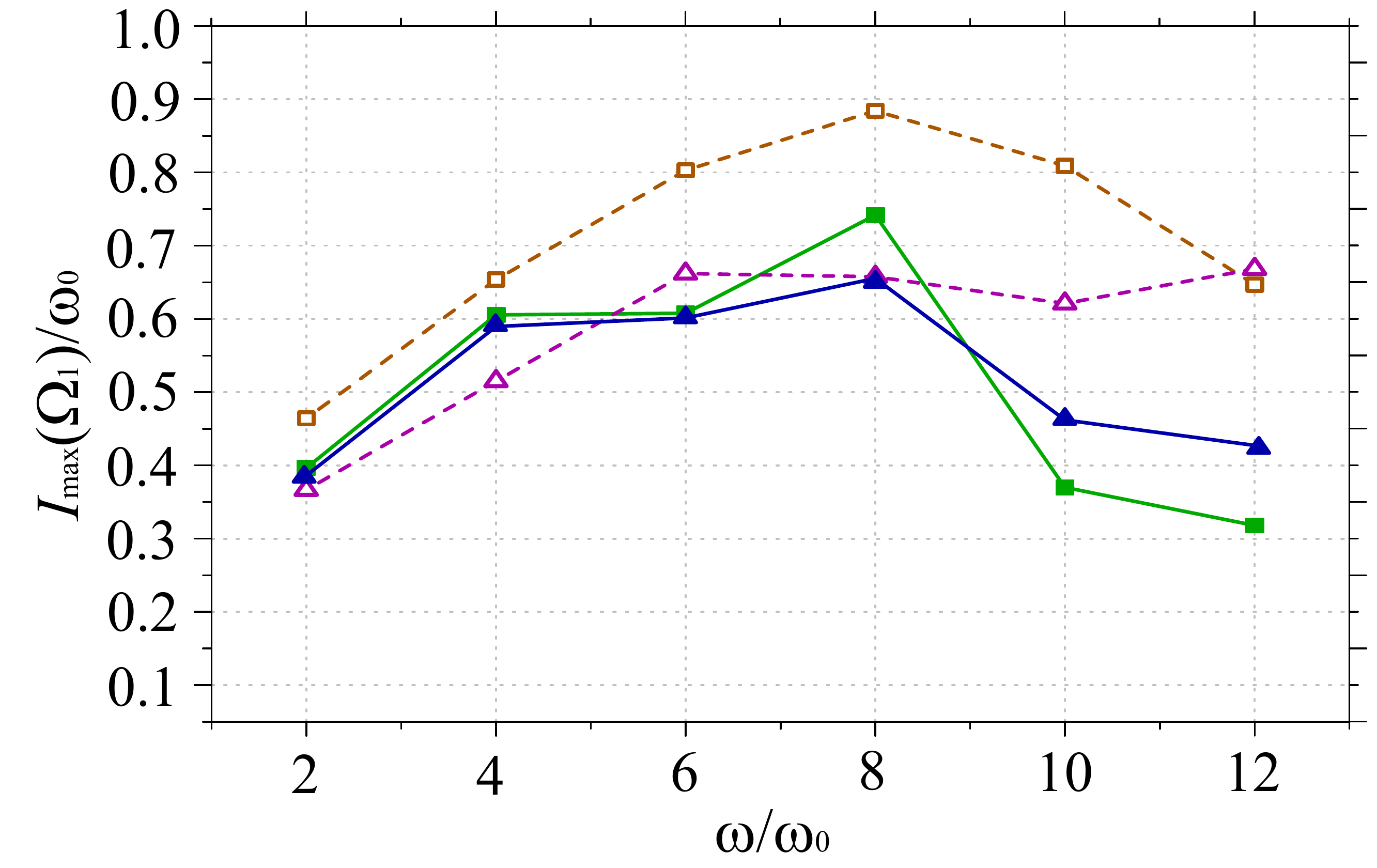}
\caption{Behavior of the global maximum of current $I_{max}(\Omega_1)$ as a function of $\omega$. Squares refer to the case with no dephasing and triangles to dephasing with $\gamma_s=\gamma_d=0.4\omega_0$ and $\gamma_1=\gamma_2=0.1\omega_0$. Empty shapes in dashed lines refer to the two-path topology and filled shapes in solid lines refer to the linear single-path topology.} 
\label{fig5}   
\end{figure}

\section{Final Remarks}
In this work, we presented a simple quantum model which takes into account the most significant features of the KcsA potassium channel. In particular, we included the fact that this system employs two pathways of conduction. From a quantum point of view, this could be a big advantage given that possible constructive interference effects can play a role. We then studied the role played by a second pathway of conduction, and we found that there might be indeed some advantage for the filter to have it. This advantage appears in both the closed system dynamics, which is certainly not the case in real ion channels, and in the open system scenario where system functions. It is important to remark that, from the experimental side, it is still necessary to wait for advances to discover whether or not this system operates in this moderate noisy regime where the two-path topology confers advantages over the single-path. In other words, it is still an open question whether or not quantum coherence is present 
in this interesting biological system. However, the measure of the current using the scheme proposed in \cite{original} would be enough to decide on the validity of the model. On the other hand, given that quantum transport is a very important topic for modern technologies, our results may still find applications in a great variety of coupled quantum systems such as arrays of quantum dots, trapped ions, and other systems alike.
%

\textit{Acknowledgments}.
A.A.C. acknowledges Funda\c c\~ao de Amparo a Pesquisa do
Estado de S\~ao Paulo (FAPESP) Grant No. 2012/12624-6, Brazil. F.L.S. acknowledge participation as member of the Brazilian National Institute of Science and Technology of Quantum Information (INCT/IQ). F.L.S. also acknowledges partial support from CNPq under grant $308948/2011-4$, Brazil. 


\end{document}